\documentclass[sigconf]{acmart}

\AtBeginDocument{%
  }

\copyrightyear{2026}
\acmYear{2026}
\setcopyright{acmlicensed}

\usepackage{multirow}
\usepackage{soul}
\usepackage{acronym}
\usepackage{amsmath}
\usepackage{graphicx}
\usepackage{makecell}
\usepackage{subcaption}
\usepackage[inline]{enumitem}

\begin{document}
\begin{sloppypar}

\acrodef{GNN}{graph neural network}
\acrodef{SR}{Sequential recommendation}
\acrodef{RNN}{recurrent neural network}
\acrodef{CNN}{convolutional neural network}
\acrodef{SSL}{self-supervised learning}
\acrodef{MC}{Markov chain}
\acrodef{GRU}{gated recurrent unit}
\acrodef{FFT}{Fast Fourier transform}
\acrodef{DL}{Deep Learning}
\acrodef{MLP}{Multilayer Perceptron}
\acrodef{CL}{contrastive learning}

\title{Quality-Aware Collaborative Multi-Positive Contrastive Learning for Sequential Recommendation}

\author{Wei Wang}
\email{iyv524896@outlook.com}
\orcid{0000-0002-7080-3381}
\affiliation{%
  \institution{Shandong University of Science and Technology}
  \streetaddress{579 Qianwangang Rd}
  \city{Qingdao}
  \country{CHINA}
  \postcode{266237}
}

\author{Yujie Lin}
\email{yu.jie.lin@outlook.com}
\orcid{0000-0002-2146-0626}
\affiliation{%
  \institution{Zhejiang Lab}
  \streetaddress{Kechuang Avenue, Zhongtai Sub-District}
  \city{Hangzhou}
  \country{CHINA}
  \postcode{311121}
}

\author{Moyan Zhang}
\email{202520529@mail.sdu.edu.cn}
\orcid{0000-0001-6130-1286}
\affiliation{%
  \institution{School of Information Science and Engineering, Shandong University}
  \streetaddress{72 Binhai Rd}
  \city{Qingdao}
  \country{CHINA}
  \postcode{266237}
}

\author{Huan Huo}
\email{Huan.Huo@uts.edu.au}
\orcid{0000-0003-2440-714X}
\affiliation{%
  \institution{Faculty of Engineering and Information Technology, University of Technology Sydney}
  \streetaddress{}
  \city{Sydney}
  \country{AUSTRALIA}
  \postcode{2007}
}

\author{Xianye Ben}
\email{benxianye@sdu.edu.cn}
\orcid{0000-0001-8083-3501}
\affiliation{%
  \institution{School of Information Science and Engineering, Shandong University}
  \streetaddress{72 Binhai Rd}
  \city{Qingdao}
  \country{CHINA}
  \postcode{266237}
}

\author{Pengjie Ren}
\email{renpengjie@sdu.edu.cn}
\orcid{0000-0003-2964-6422}
\affiliation{%
  \institution{School of Computer Science and Technology, Shandong University}
  \streetaddress{72 Binhai Rd}
  \city{Qingdao}
  \country{CHINA}
  \postcode{266237}
}

\author{Yujun Li}
\email{liyujun@sdu.edu.cn}
\orcid{0000-0003-4455-5991}
\affiliation{%
  \institution{School of Information Science and Engineering, Shandong University}
  \streetaddress{72 Binhai Rd}
  \city{Qingdao}
  \country{CHINA}
  \postcode{266237}
}

\author{Jianli Zhao}
\email{jlzhao@sdust.edu.cn}
\orcid{0000-0002-7291-9003}
\authornote{Corresponding author.}
\affiliation{%
  \institution{School of Computer Science and Engineering, Shandong University of Science and Technology}
  \streetaddress{579 Qianwangang Rd}
  \city{Qingdao}
  \country{CHINA}
  \postcode{266590}
}

\renewcommand{\shortauthors}{Wei Wang et al.}

\begin{abstract}
The effectiveness of contrastive learning (CL) in sequential recommendation hinges on the construction of contrastive views, which ideally should be both semantically consistent and diverse. 
However, most existing CL-based methods rely on heuristic augmentations that are prone to removing crucial items or disrupting transition patterns, leading to semantic drift. 
While a few studies have explored learnable augmentations to improve view quality, they often suffer from limited diversity and still necessitate heuristic aids. 
Furthermore, the quality differences across views are rarely modeled explicitly and adaptively, aggravating the false-positive issue.

To address these issues, we propose \textbf{Q}uality-aware \textbf{C}ollaborative \textbf{M}ulti-\textbf{P}ositive \textbf{C}ontrastive \textbf{L}earning for sequential recommendation (QCMP-CL). 
First, we introduce a learnable collaborative sequence augmentation module that generates two augmented views under two complementary collaborative contexts, one based on same-target sequences and the other on similar sequences, thereby enhancing view diversity while preserving intent consistency. 
Second, we design a quality-aware mechanism, tightly integrated into the model representations, which estimates each view’s quality from the confidence of its augmentation operations and assigns adaptive weights to ensure that high-confidence views contribute more supervision while low-confidence ones contribute less.
Extensive experiments on three real-world datasets demonstrate that QCMP-CL outperforms state-of-the-art CL-based sequential recommendation baselines.

\end{abstract}





\maketitle

\section{Introduction}
Sequential recommendation (SR) \cite{1-, 2-, 3-,4-} aims to model users’ evolving preferences from their interaction sequences and predict the next item, serving as a cornerstone for e-commerce and content platforms \cite{83-,84-,85-}. 
Despite the success of sequential models, sequential recommendation is still challenged by severe sparsity, long-tailed behaviors, and noisy interactions.

\begin{figure}[!ht]
  \centering
  \setlength{\abovecaptionskip}{-0.05cm}
  \setlength{\belowcaptionskip}{-0.2cm}
  \includegraphics[width=1\linewidth]{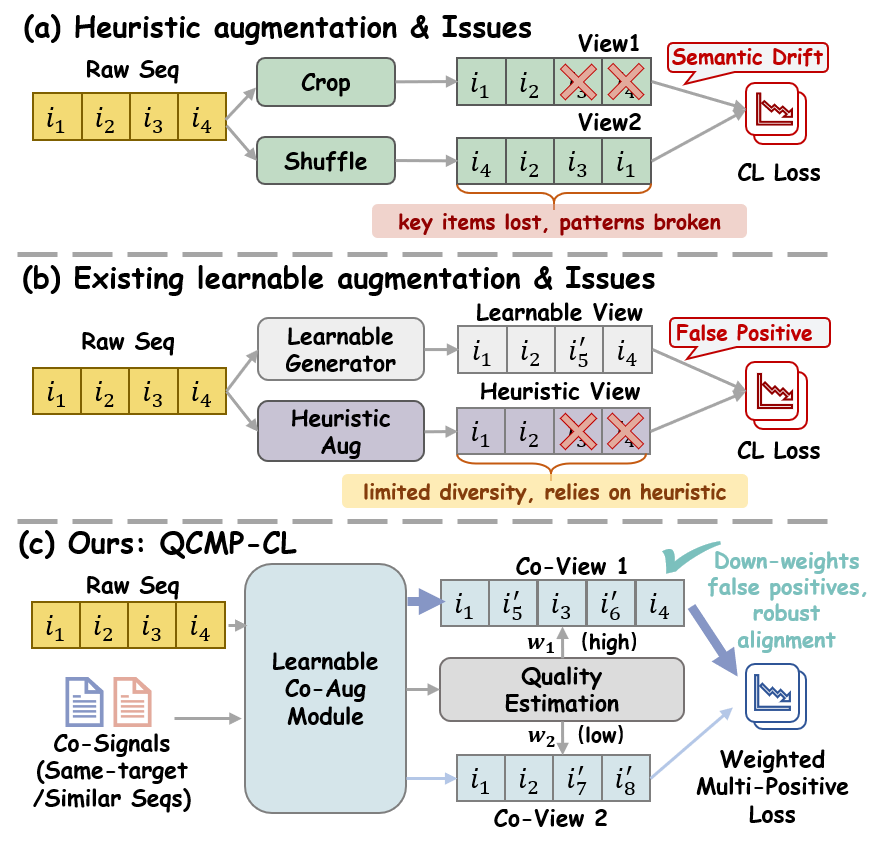}
  \caption{Illustration of contrastive view construction paradigms in sequential recommendation. (a) and (b) respectively illustrate the limitations of the heuristic methods and the existing learnable methods, (c) presents our method.}
  \Description{The figure is used to help readers understand what is presented in the introduction.}
  \label{fig1}
\end{figure}

In recent years, contrastive learning (CL) \cite{21-,22-,81-} has been extensively introduced into sequential recommendation to enhance representation learning under sparsity and noise \cite{23-,24-,25-}. 
A key idea of contrastive learning is to construct two contrastive views from the same raw sequence via data augmentation, where one view serves as the anchor and the other as a single positive sample, while augmented views from other sequences are treated as negatives. An objective such as InfoNCE \cite{17-xie,31-,80-} is then used to pull the anchor representation closer to the single positive and push it away from negative views, which encourages view-invariant interest representations. 
In general, CL-based methods \cite{26-,27-,29-} improve representation quality and downstream recommendation performance.

The quality of contrastive views directly affects the effectiveness of contrastive learning. 
Ideally, views should remain semantically consistent while introducing sufficient diversity. However, view construction in existing SR studies still has the following limitations. 
First, the constructed views do not always reflect the same semantics, or may even be completely different, which is called ``false-positive" \cite{37-,71-,72-}. 
Most methods, represented by CL4SRec \cite{17-xie}, generate views via heuristic augmentations. 
As illustrated in Figure \ref{fig1}(a), while these operations are simple and efficient, they may delete key items and distort transition patterns, causing semantic drift and false-positive issue.
Second, a few studies, exemplified by TCLARec \cite{69-TCLARec}, explore learnable augmentation to improve view quality. 
However, as illustrated in Figure \ref{fig1}(b), the generated views often lack diversity and still require heuristic operations. 
Moreover, the alignment between a view and the user’s true intent can vary sharply across different views.
Existing methods typically treat all views as equally important positives and align them with uniform force, while neglecting to explicitly model their quality differences.
This may further aggravate the false-positive issue.

To overcome the aforementioned limitations, we propose \textbf{Q}uality-aware \textbf{C}ollaborative \textbf{M}ulti-\textbf{P}ositive \textbf{C}ontrastive \textbf{L}earning for sequential recommendation (QCMP-CL). 
As illustrated in Figure \ref{fig1}(c), the key idea is to generate complementary views from collaborative signals and adaptively control the strength of supervision according to view quality, thereby improving diversity while suppressing erroneous alignments from false-positive. 
Specifically, QCMP-CL follows a two-stage pipeline. 
In stage I, we disrupt the raw sequence by randomly deleting or inserting items, and pre-train a collaborative augmentation module to recover the raw sequence by referencing collaborative sequences. 
After pre-training, the module learns to delete noisy interactions and insert auxiliary items, enabling it to construct high-quality and diverse contrastive views.
In stage II, we first provide the augmentation module with two complementary collaborative conditions --- same-target sequences and similar sequences to generate two collaborative augmented views. 
We then estimate the quality of each view based on the confidence of the module’s operation decisions, and assign adaptive quality weights, so that high-confidence views are up-weighted while low-confidence ones are down-weighted.
The underlying intuition is that more confident augmentation decisions yield more reliable views, which is verified by the ablation study and statistical analysis in Section 5.
Finally, we introduce a quality-aware multi-positive InfoNCE \cite{17-xie} objective that uses the raw sequence as the anchor and aligns it with re-weighted collaborative views. 
The objective is jointly optimized with the next-item prediction task to yield more robust and discriminative sequence representations. 
Extensive experiments on three public datasets show that QCMP-CL outperforms other state-of-the-art (SOTA) CL-based sequential recommendation models. 
To sum up, the main contributions of this work are summarized as follows:
\begin{itemize}[leftmargin=*,nosep]
    \item We propose a collaborative augmentation module that separately leverages complementary signals from same-target sequences and similar sequences to generate semantically aligned yet diverse contrastive views.
    \item We propose a confidence-based quality-aware mechanism that explicitly models view quality differences and accordingly re-weights views within contrastive learning.
    \item We conduct extensive experiments to demonstrate the effectiveness and robustness of the proposed method.
\end{itemize}


\section{Related Work}
This section discusses the view construction and quality-aware in contrastive learning-based sequential recommendation.



\subsection{View Construction in CL-Based Sequential Recommendation}
Most CL-based sequential recommendation methods construct contrastive views via data augmentation, and existing data augmentation strategies can be roughly grouped into two categories: heuristic augmentation, and model-based augmentation.

Heuristic augmentation methods perturb a raw sequence using hand-crafted rules such as masking, cropping, and reordering, where CL4SRec \cite{17-xie} is a representative work. Variants \cite{31-, 39-, 63-, 64-} that incorporate retrieved similar items/sequences or predefined edit ratios can also be viewed as heuristic in nature, since the core editing decisions are still governed by random rules. Although simple and efficient, heuristic methods may easily remove crucial items or distort transition patterns and cause semantic drift.

Many researchers have pointed out the problem of low-quality views derived from heuristic data augmentation, recent studies have explored model-based augmentation to improve the view quality. 
CCL \cite{62-} trains an attribute-enhanced data generator to fill in randomly masked positions with generated items that align with user attributes.
TCLARec \cite{69-TCLARec} further moves toward learnable view construction by learning item-wise edit operations from self-supervised restoration signals, aiming to generate augmented sequences without relying on hand-crafted rules. Despite these advances, existing model-based augmentation methods often construct limited or even only one view for the raw sequence, which limits view diversity in practice and still requires the aid of heuristic methods for contrastive learning.

\subsection{Quality-Aware Contrastive Learning}
Even with model-based view construction that improves view quality, it is still difficult to fully prevent low-quality views from introducing false positives \cite{72-,73-,74-,75-}. However, in CL-based sequential recommendation, only a few studies explicitly estimate quality differences across augmented views and control their influence during contrastive learning. SimDCL \cite{71-} points out that biased augmented samples can mislead contrastive learning and proposes a filtering mechanism to reduce such noise. TCLARec \cite{69-TCLARec} further designs a triplet contrastive objective, where learnable augmented views provide stronger supervision than heuristic views. CLEM4Rec \cite{76-} uses an external explainer to identify important items in a user’s sequence and measures view quality based on semantic similarity, then a Fourier-based filter is adopted to remove noise from sequences before contrastive learning.

In summary, existing efforts either focus on improving view construction to enhance overall view quality, or introduce coarse-grained debiasing/denoising techniques to reduce the negative impact of low-quality views. In contrast, our work not only introduces a stronger augmentation method, but also introduces an internally computed, quality-aware mechanism. This mechanism is a lightweight module tightly integrated with the model representations. It dynamically estimates the ``trustworthiness" of each augmented view based on model-intrinsic signals --- the confidence of augmentation operations. During model training, high-quality views are emphasized while low-quality ones are down-weighted. This design improves the effectiveness of CL-based sequential recommendation and helps overcome the limitations of existing studies.

\begin{figure*}[!t]
  \centering
  \includegraphics[width=\textwidth]{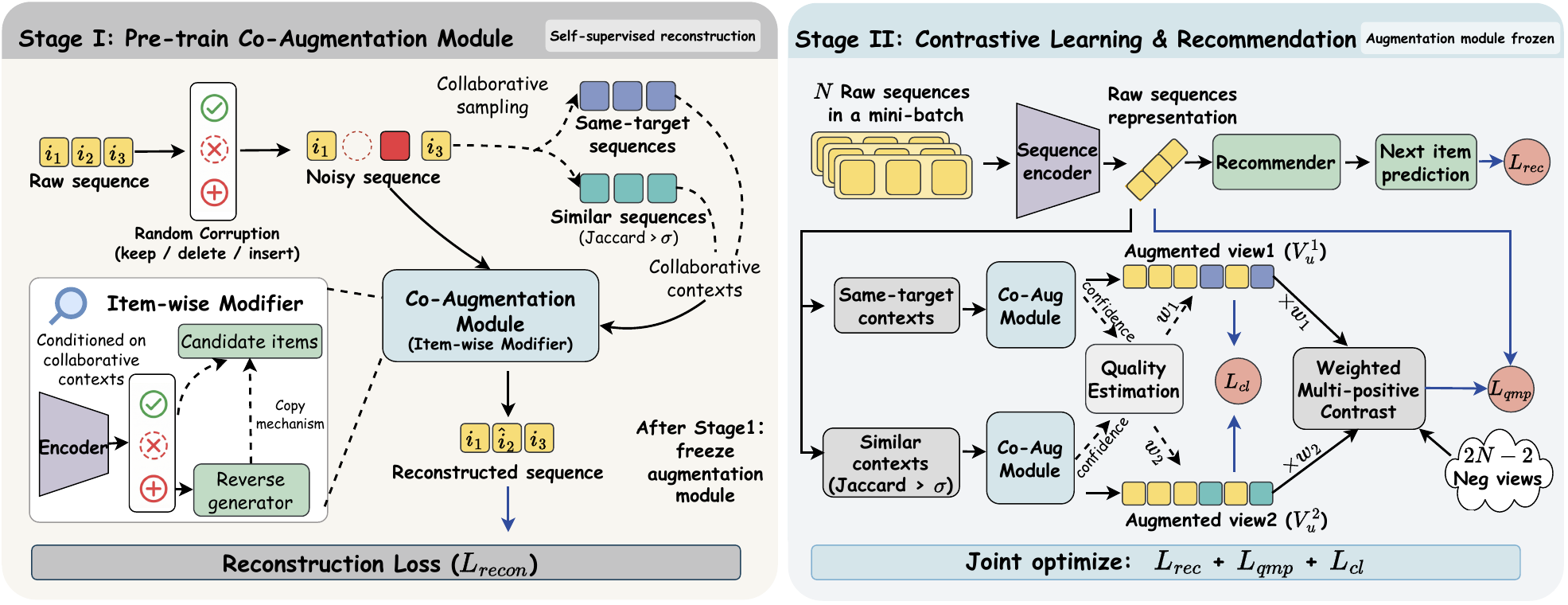}
  \caption{Overview of QCMP-CL. Stage I pre-trains the co-augmentation module via self-supervised reconstruction from randomly corrupted sequences conditioned on collaborative contexts. Stage II freezes the co-augmentation module to generate two collaborative views, computes quality weights from operation confidence, and jointly optimizes the recommender loss with contrastive objectives.}
  \Description{The figure is used to help readers understand the model.}
  \label{fig2}
\end{figure*}

\section{Method}
\subsection{Overview}
We first outline the workflow of QCMP-CL and define the main notations. 
We denote the set of users and items as $U$ and $I$. For one user, the interaction sequence is denoted as $S_u=\left[i_{1}, i_{2}, \ldots, i_{|S_u|}\right]$, where $i_{t}\in I$ represents the $t$-th item that the user interacted with. 
As shown in Figure \ref{fig2}, QCMP-CL follows a two-stage pipeline. 
We pre-train the collaborative augmentation (co-augmentation) module in stage I, and then use the pre-trained module to generate augmented views, and train the sequential recommender with a contrastive object in stage II.
During pre-training the co-augmentation module, we first generate noisy sequences by randomly applying ``keep", ``delete", or ``insert" operations to each position of the raw sequence with predefined probabilities: $[p_k,p_d,p_i]$.
Subsequently, for each noisy sequence, we sample collaborative sequences and train the augmentation module via a self-supervised reconstruction task to restore the noisy sequence to its raw sequence. 
The detailed sampling strategy for collaborative sequences is elaborated in Section 3.3.1. 
The pre-trained co-augmentation module acquires the capability to preserve critical items while deleting noise or inserting auxiliary items by referencing the collaborative contexts, thereby generating augmented views of higher quality compared to heuristic methods.

For the contrastive objective, we first sample two sets of collaborative sequences under complementary conditions for each raw sequence. 
We then feed them into the collaborative augmentation module to obtain two different augmented sequences. 
Given a mini-batch of $N$ raw sequences, we obtain $2N$ augmented sequences accordingly. 
We treat each raw sequence as the anchor, and its two augmented sequences, $V_u^{1}$ and $V_u^{2}$, as positive views. 
The remaining $2N-2$ augmented sequences in the mini-batch serve as negatives. 
To reduce the risk that false positives mislead contrastive learning, we compute view weights $w_1$ and $w_2$ from the operation confidence of the augmentation module when generating the two views.
Based on this setup, we formulate a multi-positive InfoNCE objective $L_{qmp}$ to pull the anchor closer to each positive view with different strength based on its confidence weight.
The underlying intuition is that more confident augmentation decisions yield more reliable views. 
In addition, we retain a standard InfoNCE loss $L_{cl}$ to enforce same-source consistency between $V_u^{1}$ and $V_u^{2}$ by pulling their representations closer directly, reducing view-specific drift and supplying fine-grained supervision.
Finally, the recommender is used to predict the user's next interaction. 
By jointly optimizing $L_{qmp}$, $L_{cl}$, and the recommender loss $L_{rec}$, we improve the performance of QCMP-CL.

\subsection{Sequential Recommender}
This work follows SASRec \cite{12-kang} to build a Transformer-based sequential recommender composed of an encoder and a decoder. The encoder maps the input sequences to hidden representations. We denote the item embedding matrix as $E \in \mathbb{R}^{|I| \times e}$, where $e$ is the dimension of the embedding vector, $I$ is the number of items. For the item $i_t$, the encoder indexes its embedding vector $e_{t}$ and injects positional information to obtain the initial hidden representation $h_{t}^{0}=e_{t}+p_{t}$, where $p_t$ represents the position embedding at the $t$-th position. The initial hidden representation of a sequence is formed by stacking the hidden representations of all items: $H_{e}^{0} \in \mathbb{R}^{|S_u| \times e}$.
The encoder then updates $H_{e}^{0}$ using a Transformer network:
\begin{equation}
\label{eq1}
    H_{e}^{l}=\operatorname{Trm}\left(H_{e}^{l-1}\right),
\end{equation}
where $\operatorname{Trm}$ denotes a Transformer block, $H_e^l$ denotes the representation matrix at the $l$-th layer. Finally, the encoder outputs representation of the last layer and inputs it into the augmentation module and sequential decoder. We omit the superscript and denote it as $H_e$ to simplify the notation.

The decoder is also based on a unidirectional Transformer network. Given the input sequence $S_u$ and its hidden representation matrix $H_e$, the decoder first updates $H_e$ again, takes the representation of the last layer and denotes it as $H_r$.
\begin{equation}
\label{eq2}
    H_r^0=H_e,\ H_{r}^{l}=\operatorname{Trm}\left(H_{r}^{l-1}\right).
\end{equation}
Then, the decoder calculates the probability distribution to predict the next item based on $P\left(i_{\left|S_u\right|+1} \mid S_u\right)$, which is equivalent to predict the masked item appending to the last position of $S_u$:
\begin{equation}
\label{eq3}
    P\left(i_{|S_u|+1} \mid S_u\right)=\operatorname{softmax}\left(E h_{|S_u|+1}\right),
\end{equation}
where $h_{|S_u|+1} \in \mathbb{R}^e$ denotes the hidden representation of the masked item from $H_r$.

\subsection{Model Learning}
\subsubsection{Collaborative sequences sampling}
To construct the external signals required for collaborative augmentation, we sample two types of collaborative sequences for each raw training sequence $S_u$: same-target sequences and similar sequences.

For the next-item prediction task, the last item of a sequence is treated as the target item. First, given $S_u$ and its target item $i^+$, we collect candidate sequences whose target item is also $i^+$. We sample $K$ sequences from this candidate pool as same-target collaborative sequences, denoted as $S^{col1}$. 

To obtain a more diverse yet still relevant collaborative signals, we further sample sequences that are similar to $S_u$ from the training set. Specifically, we compute the Jaccard similarity between $S_u$ and another sequence $S_v$:
\begin{equation}
\label{eq4}
    \operatorname{Jaccard}\left(S_{u}, S_{v}\right)=\frac{\left|\mathcal{I}\left(S_{u}\right) \cap \mathcal{I}\left(S_{v}\right)\right|}{\left|\mathcal{I}\left(S_{u}\right) \cup \mathcal{I}\left(S_{v}\right)\right|},
\end{equation}
where $\mathcal{I}\left(S_{u}\right)$ and $\mathcal{I}\left(S_{v}\right)$ denote the item sets of the sequences. We treat all sequences satisfying $\operatorname{Jaccard}\left(S_{u}, S_{v}\right) > \sigma$ as candidates, and sample $K$ sequences from this set as similar collaborative sequences, denoted as $S^{col2}$.

With these two sampling strategies, we build two complementary collaborative contexts for each raw sequence. $S^{col1}$ emphasizes the consistency in next-item intent, while $S^{col2}$ focuses on the overall similarity. They drive the co-augmentation module to generate diverse augmented views while preserving semantic relevance.

\subsubsection{Collaborative augmentation}
The co-augmentation module takes the collaborative sequences as external context and generates augmented sequences for each raw sequence, which are used as the views for the subsequent contrastive learning task. The module supports three operations at each position: keep the item, delete the item, or insert new items.

Given a raw sequence $S_u$ and its $K$ collaborative sequences $S^{col}$, the co-augmentation module first decides the operation at each position based on their joint representation. Specifically, we obtain the joint representation $H_e^{j}$ of the raw sequence and its collaborative sequences as follows:
\begin{equation}
\label{eq5}
    a^{k}=\operatorname{softmax}\left(H_{e} \times\left(H_{{c o l}}^{k}\right)^{T}\right), \ H_{e}^{j}=H_{e}+\left(\sum_{k=1}^{K} a^{k} \times H_{{c o l}}^{k}\right) / K,
\end{equation}
where $H_{e}$ denotes the representation of raw sequence $S_u$, $H_{c o l}^{k} \in \mathbb{R}^{\left|S^{c o l}\right| \times e}$ denotes the representation of $k$-th collaborative sequence obtained by the encoder, $a^k$ denotes the attention of $S_u$ to the $k$-th collaborative sequence.

Then we get the representation $h_t \in {\mathbb{R}^e}$ of each position from $H_e^{j}$ and calculate the probability distribution for performing the three operations:
\begin{equation}
\label{eq6}
    P\left(\hat{o}_{t} \mid S_u\right)=\operatorname{softmax}\left(W h_{t}\right),
\end{equation}
where $\hat{o}_{t}$ denotes the predicted operation, $W \in \mathbb{R}^{3 \times e}$ is a projection matrix. The operation with the highest probability will be performed. If an insertion operation is to be performed, we insert a item list of maximum length $n$ in front of the specified position through a reverse generator. Suppose the currently generated inserted list is denoted as $S_{1: n-1}^{<i_{t}}$. The generator first indexes the embedding vector $e_t$ of each item in $S_{1: n-1}^{<i_{t}}$, then stacks them with the representation $h_t$, while adding the position embedding:
\begin{equation}
\label{eq7}
    H_{c}^{0}=\left[\begin{array}{c}h_{t}+p_{1} \\e_{1}+p_{2} \\\ldots \\e_{n-1}+p_{n}\end{array}\right],
\end{equation}
where $H_c^0 \in \mathbb{R}^{n \times e}$ denotes the representation matrix of current list. We then update $H_c^0$ by Transformer again and denote the representation of the final layer as $H_c$.

Next, the generator needs to calculate the insertion probability distribution for all the candidate items. We expect items from collaborative sequences to have a greater probability of being inserted into the augmented sequence, so that collaborative sequences will provide a stronger supervision signal in contrastive learning. Specifically, we apply a copy mechanism \cite{77-Ren,78-CLOUD}, which first obtains the attention score $e_k^n$ for each collaborative sequences:
\begin{equation}
\label{eq8}
    e_{k}^{n}=v_{c o}^{T} \tanh \left(W_{c o} h_{k}+U_{c o} h_{n}\right),
\end{equation}
where $h_n$ denotes the representation of the last position of $H_c$, $W_{c o} \in \mathbb{R}^{e \times e}$, $U_{c o} \in \mathbb{R}^{e \times e}$ and $v_{c o} \in \mathbb{R}^{e \times 1}$ are transition matrices, $h_k$ denotes the average representation of items in $k$-th collaborative sequence. Then the context representation $c^n$ from all collaborative sequences is obtained by weighted sum:
\begin{equation}
\label{eq9}
    a_{k}^{n}=\operatorname{softmax}\left(e_{k}^{n}\right), \ c^{n}=\sum_{k=1}^{K} a_{k}^{n} h_{k}.
\end{equation}
We follow Eq.\ref{eq10} to calculate the probability distribution of inserting items from collaborative sequences or the full candidates:
\begin{equation}
\label{eq10}
    \left[P\left(I^{c o l}\right), P\left(I^{all}\right)\right] =\operatorname{softmax}\left(W_{c o}^{P} c^{n}+W_{a l l}^{P} h_{n}\right),
\end{equation}
where $W_{co}^P\in\mathbb{R}^{2 \times e}$ and $W_{all}^P \in \mathbb{R}^{2 \times e}$ are transition matrices, $I^{col}$ denotes the items from collaborative sequences and $I^{all}$ denotes all candidates. Finally, we follow Eq.\ref{eq11} - Eq.\ref{eq12} to calculate the probability of the next inserted item $i_j$:
\begin{equation}
\label{eq11}
    P_{c o l}\left(i_{j}\right)=\frac{\exp \left(e_{j}^T h_{n}\right) \times N_j}{\sum_{j}^{|I|} \exp \left(e_{j}^T h_{n}\right) \times N_j}, \ 
     P_{a l l}\left(i_{j}\right)=\operatorname{softmax}\left(e_{j}^T h_{n}\right),
\end{equation}
\begin{equation}
\label{eq12}
    P\left({i}_{j} \mid S_{1: n-1}^{<i_{t}}\right)=P\left(I^{c o l}\right) \times P_{c o l}\left(i_{j}\right)+P\left(I^{all}\right) \times P_{a l l}\left(i_{j}\right),
\end{equation}
where $N_j$ indicates how many times $i_j$ occurs in collaborative sequences, and $e_j \in \mathbb{R}^e$ denotes the embedding vector of $i_j$.
Through the above calculation, an item’s insertion probability will grow with its occurrence count in collaborative sequences.

\subsubsection{Quality-aware multi-positive contrastive learning}
\label{3.3.3}
Feeding the same-target collaborative sequences and the similar collaborative sequences into the pre-trained co-augmentation module yields two different augmented views. Given a mini-batch of size $N$, the co-augmentation module generates $2N$ augmented views. We treat each raw sequence $S_u$ as the anchor, and its two augmented views as a positive pair $[V_u^{1}, V_u^{2}]$, the remaining $2N-2$ augmented views in the mini-batch (from other raw sequences) are used as negatives. We then apply a multi-positive InfoNCE objective to align $S_u$ with its positives. Specifically, we follow Eq.\ref{eq13} to increase the representation similarity between $S_u$ and both $V_u^{1}$ and $ V_u^{2}$:
\begin{equation}
\label{eq13}
    L_{qmp}=-\sum_{k \in\{1,2\}} w_{k} \log \frac{\exp \left(s\left(S_{u}, V_{u}^{k}\right)\right)}{\exp \left(s\left(S_{u}, V_{u}^{k}\right)\right)+\sum_{\mathcal{N}(u)} \exp \left(s\left(S_{u}, \mathcal{N}(u)\right)\right)},
\end{equation}
where $\mathcal{N}(u)=\left\{V_{v}^{k} \mid v \neq u, k \in\{1,2\}\right\}$ denotes the negative set, $s(\cdot)$ is the dot-product similarity.

In Eq.\ref{eq13}, we introduce a quality weight $w_k$ for each positive view to control its supervision strength in $L_{qmp}$. This is motivated by the fact that the quality between positive views is not exactly equivalent, and such differences should be modeled explicitly to further avoid false-positive alignments caused by low-quality views. The quality weights are computed from the operation confidence of the co-augmentation module during view construction. At each position, the module outputs an operation distribution $P\left(\hat{o}_{t} \mid S_u\right)$ over keep/delete/insert. We use the maximum probability as the confidence for that position, average it over all non-padding positions to obtain a sample-level quality score for the view, and then apply a softmax to get normalized weights:
\begin{equation}
    q_{k}=\frac{1}{|S_u|}\sum_{t=1}^{|S_u|}\max_{o\in\{\text{k,d,i}\}}P\left(\hat{o}_{t} = o \mid S_u\right),
\end{equation}
\begin{equation}
\label{eq15}
    [w_{1},w_{2}]=\mathrm{softmax}\left([q_1,q_2]/q_{temp}\right),
\end{equation}
where $q_{temp}$ is a temperature parameter. The intuition is that more confident augmentation decisions generate more reliable views, which should therefore contribute stronger supervision in contrastive learning.

Finally, we retain the standard InfoNCE objective as an auxiliary term that provides weaker supervision, preventing the two collaborative views from the same source from drifting too far apart in the representation space:
\begin{equation}
\label{eq16}
    L_{cl}=-\log \frac{\exp \left(s\left(V_{u}^{1}, V_{u}^{2}\right)\right)}{\exp \left(s\left(V_{u}^{1}, V_{u}^{2}\right)\right)+\sum_{\mathcal{N}(u)} \exp \left(s\left(V_{u}^{1}, \mathcal{N}(u)\right)\right)}.
\end{equation}

\subsubsection{Model training}
The training procedure follows a two-stage pipeline. In Stage I, we pre-train the co-augmentation module via random corruption and a reconstruction task. Specifically, we perturb each raw sequence using a random operation distribution (e.g., $[0.4,0.5,0.1]$) for keep, delete, and insert, the corrupted sequence is denoted as $S^m$. The reconstruction task then requires the module to remove the inserted noisy items and recover the original items that were deleted, with the operations denoted as $O$. This process is applied iteratively to each raw sequence, allowing the pre-trained augmentation module to learn to eliminate noise and insert highly relevant auxiliary items. The target loss function is to minimize the negative log-likelihood of the probability $P\left(S \mid S^{m}\right)$:
\begin{equation}
\label{eq17}
\resizebox{1.0\hsize}{!}{$
    \begin{aligned}L_{recon} & =-\log P\left(S \mid S^{m}\right) \\& =-\left(\log P\left(O \mid S^{m}\right)+\sum_{i \in I^{i n s}} \log P\left(S^{<i} \mid S^{m}\right)\right) \\& =-\left(\sum_{t=1}^{\left|S^{m}\right|} \log P\left(\hat{o}_{t}=o_{t} \mid S^{m}\right)+\sum_{i \in I^{i n s}} \sum_{n=1}^{\left|S^{<i}\right|} \log P\left(\hat{i}_{n}=i_{n} \mid S_{1: n-1}^{<i}, S^{m}\right)\right)\end{aligned}$},
\end{equation}
where $I^{ins}$ indicates the positions where items need to be inserted, $S^{<i}$ represents the ground truth of the inserted items.

In stage II, the recommender is optimized by minimizing the negative log-likelihood of the probability $P\left(\hat{i}_{|S_u|} \mid S_u\right)$:
\begin{equation}
    \begin{aligned}
        L_{rec} &=-\log P\left(\hat{i}_{|S_u|} \mid S_u\right) \\& =-\log P\left(\hat{i}_{|S_u|}= i_{|S_u|} \mid S_u \right),
    \end{aligned}
\end{equation}
where $i_{|S_u|}$ is the masked item in $S_u$.

At last, the standard backpropagation algorithm \cite{30-,79-} is adopted to minimize the weighted joint loss of the recommender and the contrastive learning objects:
\begin{equation}
    L_{joint}=L_{rec}+\alpha L_{qmp}+\beta L_{cl},
\end{equation}
where \(\alpha\) and \(\beta\) are weight hyperparameters. In practice, $\beta$ is set smaller (e.g., 0.05) to avoid over-constraining the two augmented views, while $L_{qmp}$ (e.g., $\alpha=0.2$) serves as the main contrastive supervision to improve robustness and representation quality.
\section{EXPERIMENTAL SETUP}

\subsection{Datasets}
\begin{table}[h]
    \setlength{\aboverulesep}{0pt}
    \setlength{\belowrulesep}{0pt}
	\centering
	\caption{Statistics of the datasets.}
	\label{tab1}
	\begin{tabular}{l|c|c|c|c|c}
		\toprule
		Datasets & Users & Items & Records &Avg.len &Density\\ 
		\midrule 
        Beauty &22,362 &12,101 &194,682 &8.7 &0.07\% \\
        Yelp &22,844 &16,552 &236,999 &10.4 &0.06\% \\
        Sports &35,597 &18,357 &294,483 &8.3 &0.05\% \\
        \bottomrule 
	\end{tabular}
\end{table}

\noindent For a fair comparison with other SOTA baselines \cite{30-,69-TCLARec}, we evaluate the performance of the models on three public datasets: Beauty, Yelp, and Sports. 
Beauty and Sports are two product review datasets crawled from Amazon \cite{59-}. Yelp is a business recommendation dataset released by Yelp.com. We also follow \cite{30-,69-TCLARec} to preprocess the datasets. 
First, we filtered out users and items with fewer than 5 interaction records. 
Then, we get the item sequence by sorting the interaction records in order. The statistics of the preprocessed datasets are shown in Table \ref{tab1}.

\subsection{Baselines}
We compare QCMP-CL with eight SOTA sequential recommendation baselines and classify them into three groups: vanilla models, contrastive learning-based models and correction-based model.

\begin{itemize}
    \item \textbf{Vanilla recommendation models:}
    \item[-] \textbf{SASRec} \cite{12-kang} introduces the Transformer module to model the transition relationship between items.
    \item[-] \textbf{BERT4Rec} \cite{13-sun} introduces a bidirectional Transformer and trained by masked item prediction task.
    \item \textbf{Contrastive learning-based models:}
    \item[-] \textbf{CL4SRec} \cite{17-xie} is a representative of the heuristic CL-based sequential recommendation, which combines three random sequence augmentation methods.
    \item[-] \textbf{DuoRec} \cite{58-} employs an augmentation method based on dropout and a novel sampling strategy to construct contrastive self-supervised signals.
    \item[-] \textbf{CoSeRec} \cite{31-} designs two SSL-based sequence augmentation methods and combines them with random operations.
    \item[-] \textbf{ICSRec} \cite{39-} segments user intentions from the raw sequence and introduces intent contrastive learning.
    \item[-] \textbf{TCLARec} \cite{69-TCLARec} proposes a learnable sequence augmentation module, and combined with another heuristic augmented sequence to perform contrastive learning.
    \item \textbf{Sequence correction-based model:}
    \item[-] \textbf{STEAM} \cite{30-} trains a corrector through a self-supervised task to remove noisy items in native sequences and insert missed items, aiming to improve recommendation accuracy.
\end{itemize}

\begin{table*}[!ht]
    \setlength{\tabcolsep}{4pt}
    \renewcommand{\arraystretch}{0.8}
    \caption{Overall performance comparison of different methods on the three datasets. The best performance and the second best performance are denoted in bold and underlined fonts respectively. * denotes the improvement is significant at $p$ < 0.05.}
    \label{tab2}
    \scalebox{0.95}{
    \begin{tabular}{c | c | cc ccc c c c c | c | c}
      \toprule
      \multirow{2}{*}{Dataset} & \multirow{2}{*}{Metrics} & \multirow{2}{*}{SASRec} & \multirow{2}{*}{BERT4Rec} & \multirow{2}{*}{CL4SRec} & \multirow{2}{*}{DuoRec} & \multirow{2}{*}{CoSeRec} & \multirow{2}{*}{ICSRec} & \multirow{2}{*}{STEAM} & \multirow{2}{*}{TCLARec} & \multirow{2}{*}{QCMP-CL} & \multicolumn{2}{c}{improve v.s.} \\ \cline{12-13} 
      & & & & & & & & & & & \multicolumn{1}{c|}{CL4SRec} & All \\ 
      \midrule

      \multirow{10}{*}{Beauty} &HR@5 &0.3721 &0.3666 &0.4067 &0.4094 &0.4124 &0.4250 &0.4284 &\underline{0.4446} &\textbf{0.4590*} &12.85\% &3.23\% \\
      &HR@10 &0.4639 &0.4728 &0.5056 &0.5187 &0.5087 &0.5188 &0.5256 &\underline{0.5460} &\textbf{0.5632*} &11.39\% &3.15\% \\
      &HR@20 &0.5804 &0.6011 &0.6199 &0.6200 &0.6356 &0.6345 &0.6486 &\underline{0.6606} &\textbf{0.6755*} &8.96\% &2.25\% \\
      &MRR@5 &0.2611 &0.2337 &0.2785 &0.2884  &0.2749 &0.2963 &0.2899 &\underline{0.3161} &\textbf{0.3252*} &16.76\% &2.87\% \\
      &MRR@10 &0.2732 &0.2478 &0.2916 &0.3015 &0.2890 &0.3087 &0.3028 &\underline{0.3297} &\textbf{0.3391*} &16.28\% &2.85\% \\
      &MRR@20 &0.2812 &0.2566 &0.2994 &0.3092 &0.2971 &0.3167 &0.3112 &\underline{0.3376} &\textbf{0.3468*} &15.83\% &2.72\% \\
      &NDCG@5 &0.2887 &0.2570 &0.3104 &0.3194 &0.3090 &0.3284 &0.3181 &\underline{0.3482} &\textbf{0.3585*} &15.49\% &2.95\% \\
      &NDCG@10 &0.3182 &0.2907 &0.3422 &0.3494 &0.3434 &0.3586 &0.3508 &\underline{0.3810} &\textbf{0.3922*} &14.61\% &2.93\% \\
      &NDCG@20 &0.3476 &0.3227 &0.3710 &0.3774 &0.3728 &0.3878 &0.3807 &\underline{0.4099} &\textbf{0.4205*} &13.34\% &2.58\% \\
      
      \midrule
      \multirow{10}{*}{Yelp} &HR@5 &0.5847 &0.6118 &0.6292 &0.6400 &0.6454 &0.6618 &0.6683 &\underline{0.6833} &\textbf{0.7033*} &11.77\% &2.92\% \\
      &HR@10 &0.7833 &0.7972 &0.8223 &0.8263 &0.8241 &0.8337 &0.8460 &\underline{0.8569} &\textbf{0.8698*} &5.77\% &1.50\% \\
      &HR@20 &0.9194 &0.9235 &0.9486 &\underline{0.9564} &0.9444 &0.9531 &0.9511 &0.9554 &\textbf{0.9582} &1.01\% &0.19\% \\
      &MRR@5 &0.3543 &0.3764 &0.3991 &0.4085 &0.4140 &0.4302 &0.4293 &\underline{0.4482} &\textbf{0.4619*} &15.73\% &3.05\% \\
      &MRR@10 &0.3808 &0.4013 &0.4251 &0.4334 &0.4380 &0.4533 &0.4531 &\underline{0.4716} &\textbf{0.4845*} &13.97\% &2.73\% \\
      &MRR@20 &0.3907 &0.4104 &0.4343 &0.4429 &0.4467 &0.4619 &0.4607 &\underline{0.4788} &\textbf{0.4910*} &13.05\% &2.54\% \\
      &NDCG@5 &0.4113 &0.4257 &0.4562 &0.4647 &0.4714 &0.4877 &0.4895 &\underline{0.5066} &\textbf{0.5219*} &14.86\% &3.02\% \\
      &NDCG@10 &0.4756 &0.4870 &0.5188 &0.5245 &0.5293 &0.5434 &0.5461 &\underline{0.5630} &\textbf{0.5761*} &11.04\% &2.32\% \\
      &NDCG@20 &0.5105 &0.5218 &0.5513 &0.5590 &0.5602 &0.5741 &0.5735 &\underline{0.5883} &\textbf{0.5990*} &8.65\% &1.81\% \\
      
      \midrule
      \multirow{10}{*}{Sports} &HR@5 &0.3450 &0.3515 &0.3935 &0.3980 &0.4074 &0.4220 &0.4190 &\underline{0.4379} &\textbf{0.4547*} &15.55\% &3.83\% \\
      &HR@10 &0.4597 &0.4791 &0.5152 &0.5192 &0.5305 &0.5459 &0.5496 &\underline{0.5645} &\textbf{0.5813*} &12.82\% &2.97\% \\
      &HR@20 &0.5942 &0.6280 &0.6562 &0.6583 &0.6650 &0.6861 &0.6916 &\underline{0.7017} &\textbf{0.7182*} &9.44\% &2.35\% \\
      &MRR@5 &0.2204 &0.2154 &0.2600 &0.2597 &0.2653 &0.2741 &0.2685 &\underline{0.2896} &\textbf{0.2992*} &15.07\% &3.31\% \\
      &MRR@10 &0.2357 &0.2323 &0.2762 &0.2758 &0.2817 &0.2906 &0.2858 &\underline{0.3064} &\textbf{0.3160*} &14.40\% &3.13\% \\
      &MRR@20 &0.2449 &0.2426 &0.2859 &0.2854 &0.2910 &0.3003 &0.2956 &\underline{0.3159} &\textbf{0.3255*} &13.85\% &3.03\% \\
      &NDCG@5 &0.2513 &0.2448 &0.2931 &0.2936 &0.3005 &0.3108 &0.3058 &\underline{0.3264} &\textbf{0.3378*} &15.25\% &3.49\% \\
      &NDCG@10 &0.2883 &0.2862 &0.3324 &0.3335 &0.3403 &0.3508 &0.3479 &\underline{0.3673} &\textbf{0.3787*} &13.92\% &3.10\% \\
      &NDCG@20 &0.3222 &0.3236 &0.3680 &0.3686 &0.3743 &0.3862 &0.3838 &\underline{0.4020} &\textbf{0.4133*} &12.30\% &2.81\% \\
      \bottomrule
    \end{tabular}
    }
\end{table*}

\subsection{Metrics and Implementation Details}
We adopt HR, MRR, and NDCG as metrics to evaluate the performance of all models, the length of recommendation list is $\{5,10,20\}$. For each item sequence, the last item is the test item, the second last item is the validation item, and the remaining items are used for training. We limit the maximum length of the raw sequence to 50. We follow \cite{13-sun, 30-, 69-TCLARec} to randomly select 99 uninteracted items as negative samples for each validation and test item to evaluate the recommendation performance. For all models, we initialize the model parameters using the Xavier method \cite{61-} and train the models using the Adam optimizer \cite{82-}. The learning rate is set to 0.001 and embedding size is set to 64. 

For QCMP-CL, the number of heads and the layers in Transformer is set to 2, the dropout rate is 0.5. The mini-batch size is 256 on Beauty and Yelp, and 128 on Sports. The co-augmentation module can insert up to 5 items consecutively at each position, and the maximum length of the augmented sequence is limited to 60. The similarity threshold $\sigma$ is set to 0.2. In stage I, we sample 30 same-target sequences and 30 similar sequences for each raw sequence as collaborative contexts to train the co-augmentation module. In stage II, we sample 10 same-target sequences and 10 similar sequences to construct contrastive views.

\section{EXPERIMENTAL RESULTS}
\subsection{Overall Performance}

We first compare the overall performance of QCMP-CL with eight baseline models on three datasets, the experimental results are shown in Table \ref{tab2}. From the experimental results, we have the following observations. 

Overall, QCMP-CL achieves the best results on all metrics, delivering stable and  significant improvements over CL4SRec and other strong CL-based sequential recommendation methods. Specifically, the baselines reveal an evident performance progression. SASRec and BERT4Rec, as pure self-supervised SR models, are consistently weaker than CL-based methods, indicating that contrastive learning provides additional effective supervision under sparse and noisy interactions. CL4SRec brings a basic improvement by constructing views via random perturbations, but its heuristic augmentations may distort transition patterns, leading to limited gains. DuoRec and CoSeRec further improve performance by constructing more natural positives or incorporating retrieval signals, surpassing purely random augmentations, yet they are still constrained by heuristic rules such as random edit position/scale. ICSRec and STEAM validate the importance of view quality through more refined view construction or correction mechanisms. On top of these, as a representative learnable augmentation method, TCLARec typically achieves the strongest baseline performance, suggesting that learnable sequence augmentation can generate higher-quality views than heuristic perturbations. However, it mainly relies on a single augmented view and does not explicitly model quality differences of views, leaving room for further improvement.

In contrast, QCMP-CL maintains top performance stably across all three datasets. Compared with CL4SRec, QCMP-CL obtains relative improvements of about 8\%–16\% on Beauty, 1\%–15\% on Yelp, and 9\%–15\% on Sports. Compared with the strongest baseline, QCMP-CL still achieves steady gains of about 2.3\%–3.8\% on Beauty and Sports, while the improvement on Yelp is smaller, which aligns with the expected diminishing marginal improvement once recall approaches saturation. These improvements mainly come from two aspects. First, collaborative multi-view construction leverages two complementary collaborative conditions to construct two highly relevant augmented views for each raw sequence, mitigating the diversity limitation of single-view augmentation. Second, quality-aware multi-positive alignment explicitly models quality differences among positive views. It assigns adaptive weights based on the augmentation confidence, strengthening supervision from high-quality views while down-weighting low-quality ones, thereby reducing the risk of false-positive alignments. Meanwhile, the auxiliary standard  InfoNCE term constrains same-source view drift and improves training stability. Taken together, the results suggest that the ``collaborative multi-view + quality weight" design in sequential recommendation can better exploit contrastive supervision than heuristic augmentation or single-view learnable augmentation, leading to stronger recommendation performance.

\subsection{Ablation Study}
\label{5.2}
To directly verify the effectiveness of each component in QCMP-CL, we design five variants and compare their performance:
\begin{itemize}
    \item \textbf{$w/o\text{-}L_{cl}$}: removes the standard InfoNCE contrastive loss that constrains representation drift between same-source views;
    \item \textbf{$w/o\text{-}Qa$}: removes the quality weights and aligns the raw sequence with the two positive views with equal strength;
    \item \textbf{$w\text{-}col1$}: keeps the same-target collaborative view, while the other view is generated by random augmentation;
    \item \textbf{$w\text{-}col2$}: keeps the similar collaborative view, while the other view is generated by random augmentation;
    \item Heuristic: generates both views via random augmentation.
\end{itemize}
$w/o\text{-}L_{cl}$ and $w/o\text{-}Qa$ are used to examine the effectiveness of our contrastive objectives, while $w\text{-}col1$, $w\text{-}col2$, and Heuristic mainly compare different view construction strategies.

The results are reported in Table \ref{tab3}. First, removing the auxiliary contrastive loss $L_{cl}$ leads to a slight overall performance drop. This is because the main supervision in QCMP-CL comes from the multi-positive contrastive objective $L_{qmp}$. $L_{cl}$ acts more like a regularizer that mitigates representation drift between the two same-source views, with gains mainly reflected in ranking quality (MRR/NDCG) and training stability. 
In contrast, $w/o\text{-}Qa$ exhibits a more pronounced degradation, which aligns with our intuition: the two collaborative views are not equally reliable, and enforcing equal-strength alignment amplifies the risk of false-positive alignment, damaging fine-grained ranking.

Next, when only one collaborative view is retained and paired with a randomly augmented view, both $w\text{-}col1$ and $w\text{-}col2$ perform substantially worse than full QCMP-CL, with an overall drop of about 3.7\%--6.9\%. In fact, $w\text{-}col1$ and $w\text{-}col2$ are highly similar to TCLARec \cite{69-TCLARec}, but they lack the triplet contrastive constraint, which explains why they under-perform TCLARec. This indicates that relying on a single collaborative view is insufficient to provide strong supervision, while random augmentation introduces additional noise and weakens the contrastive signal. We further observe that $w\text{-}col1$ consistently and slightly outperforms $w\text{-}col2$ on all three datasets. We attribute this to the fact that the same-target condition is more directly aligned with the next-item intent and typically provides more accurate collaborative signals. In comparison, the similarity-based condition covers a broader scope but is also more likely to introduce semantic shift, making it more vulnerable when quality weighting and multi-view complementarity are absent. Finally, the Heuristic variant essentially degenerates to CL4SRec, and its performance drops by about 8.37\%--10.94\% relative to the full model. This again confirms that random perturbations can easily disrupt critical transition patterns and fail to provide stable, high-quality positive supervision.

In summary, collaborative sequence augmentation and the adaptive quality-aware mechanism are the key contributors to QCMP-CL, while $L_{cl}$ provides an additional same-source consistency regularization that further stabilizes performance.

\begin{table*}[ht]
    \renewcommand{\arraystretch}{0.8}
    \caption{Ablation study. The best performance is denoted in bold fonts. Sum is the total of the HR, MRR, and NDCG.}
    \label{tab3}
    \scalebox{0.9}{
    \begin{tabular}{c|c| ccc ccc ccc c}
         \toprule
         Dataset & Variants & HR@5 & HR@10 & HR@20 & MRR@5 & MRR@10 & MRR@20 & NDCG@5 & NDCG@10 & NDCG@20 & Sum \\
         \midrule
         \multirow{6}{*}{Beauty} 
         & QCMP-CL &\textbf{0.4590} &\textbf{0.5632} &\textbf{0.6755} &\textbf{0.3252} &\textbf{0.3391} &\textbf{0.3468} &\textbf{0.3585} &\textbf{0.3922} &\textbf{0.4205} &\textbf{3.8803} \\
         &$w/o\text{-}L_{cl}$ &0.4546 &0.5604 &0.6747 &0.3214 &0.3355 &0.3434 &0.3545 &0.3888 &0.4176 &3.8514  \\
         &$w/o\text{-}Qa$ &0.4519 &0.5570 &0.6713 &0.3136 &0.3277 &0.3356 &0.3480 &0.3821 &0.4109 &3.7984  \\
         &$w\text{-}col1$ &0.4422 &0.5435 &0.6597 &0.3029 &0.3164 &0.3245 &0.3376 &0.3703 &0.3997 &3.6972\\
         &$w\text{-}col2$ &0.4393 &0.5423 &0.6601 &0.2995 &0.3133 &0.3215 &0.3343 &0.3677 &0.3974 &3.6758 \\
         &Heuristic &0.4136 &0.5163 &0.6318 &0.2774 &0.2911 &0.2991 &0.3113 &0.3445 &0.3737 &3.4594 \\

         \midrule
         \multirow{6}{*}{Yelp} 
         & QCMP-CL &\textbf{0.7033} &0.8698 &0.9582 &\textbf{0.4619} &\textbf{0.4845} &\textbf{0.4910} &\textbf{0.5219} &\textbf{0.5761} &\textbf{0.5990} &\textbf{5.6661}  \\
         &$w/o\text{-}L_{cl}$ &0.7007 &\textbf{0.8716} &0.9629 &0.4561 &0.4792 &0.4860 &0.5169 &0.5725 &0.5961 &5.6425 \\
         &$w/o\text{-}Qa$ &0.6901 &0.8668 &\textbf{0.9644} &0.4446 &0.4685 &0.4757 &0.5056 &0.5631 &0.5883 &5.5675 \\
         &$w\text{-}col1$ &0.6763 &0.8597 &0.9643 &0.4297 &0.4526 &0.4604 &0.4896 &0.5491 &0.5762 &5.4565 \\
         &$w\text{-}col2$ &0.6729 &0.8572 &0.9640 &0.4263 &0.4511 &0.4590 &0.4875 &0.5473 &0.5750 &5.4406 \\
         &Heuristic &0.6286 &0.8238 &0.9581 &0.3976 &0.4238 &0.4336 &0.4549 &0.5182 &0.5528 &5.1919 \\

         \midrule
         \multirow{6}{*}{Sports} 
         & QCMP-CL &\textbf{0.4547} &\textbf{0.5813} &\textbf{0.7182} &\textbf{0.2992} &\textbf{0.3160} &\textbf{0.3255} &\textbf{0.3378} &\textbf{0.3787} &\textbf{0.4133} &\textbf{3.8251} \\
         &$w/o\text{-}L_{cl}$ &0.4498 &0.5778 &0.7137 &0.2965 &0.3136 &0.3231 &0.3345 &0.3760 &0.4104 &3.7957 \\
         &$w/o\text{-}Qa$ &0.4429 &0.5732 &0.7136 &0.2883 &0.3056 &0.3154 &0.3267 &0.3687 &0.4043 &3.7391 \\
         &$w\text{-}col1$ &0.4234 &0.5536 &0.6958 &0.2718 &0.2892 &0.2990 &0.3095 &0.3515 &0.3874 &3.5816 \\
         &$w\text{-}col2$ &0.4206 &0.5506 &0.6929 &0.2700 &0.2873 &0.2972 &0.3074 &0.3494 &0.3854 &3.5611 \\ 
         &Heuristic &0.4022 &0.5322 &0.6755 &0.2535 &0.2708 &0.2807 &0.2904 &0.3324 &0.3686 &3.4066 \\

         \bottomrule
    \end{tabular}
    }
\end{table*}

\subsection{Statistical Analysis}
\label{5.3}

We conduct a statistical analysis to support the claim that ``high-confidence views are generally of higher quality." 
Specifically, we adopt an intuitive proxy to measure view quality. 
We define an item and the target item as ``co-occurrence" if they appear in the same sequence in the dataset. 
Because the items that naturally co-occur with the target item are more semantically relevant, we treat the average co-occurrence count over all items in an augmented view as its quality score.
We then report the proportion of cases where the high-confidence view achieves a higher quality score than the low-confidence view.

\begin{figure}[t]
   \centering
   \begin{minipage}[t]{.485\linewidth}
      \includegraphics[width=\linewidth]{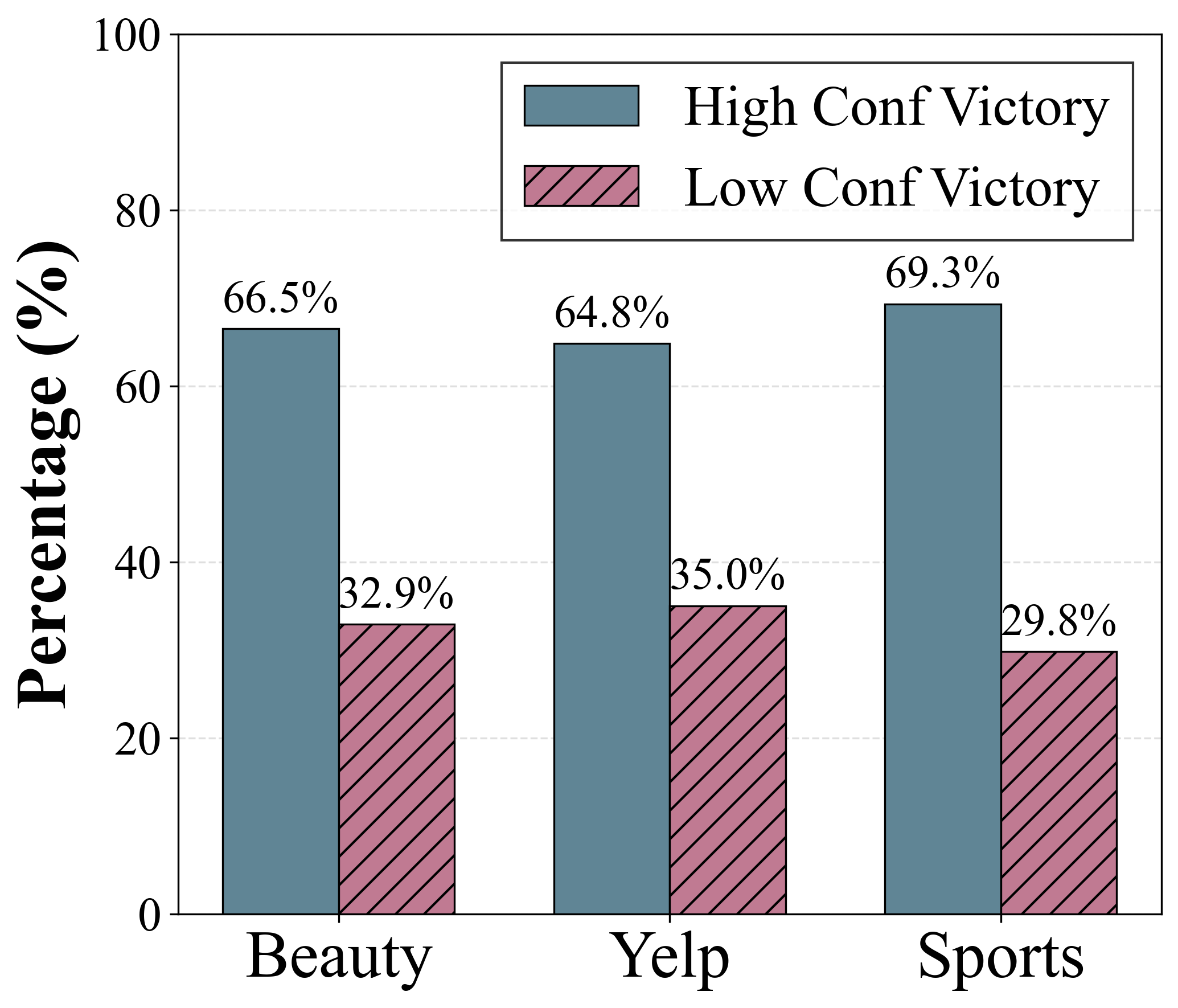}
      \caption{Statistical Analysis on Beauty, Yelp and Sports. }
      \label{fig 3}
   \end{minipage}
   \hfill
   \begin{minipage}[t]{.485\linewidth}
      \includegraphics[width=\linewidth]{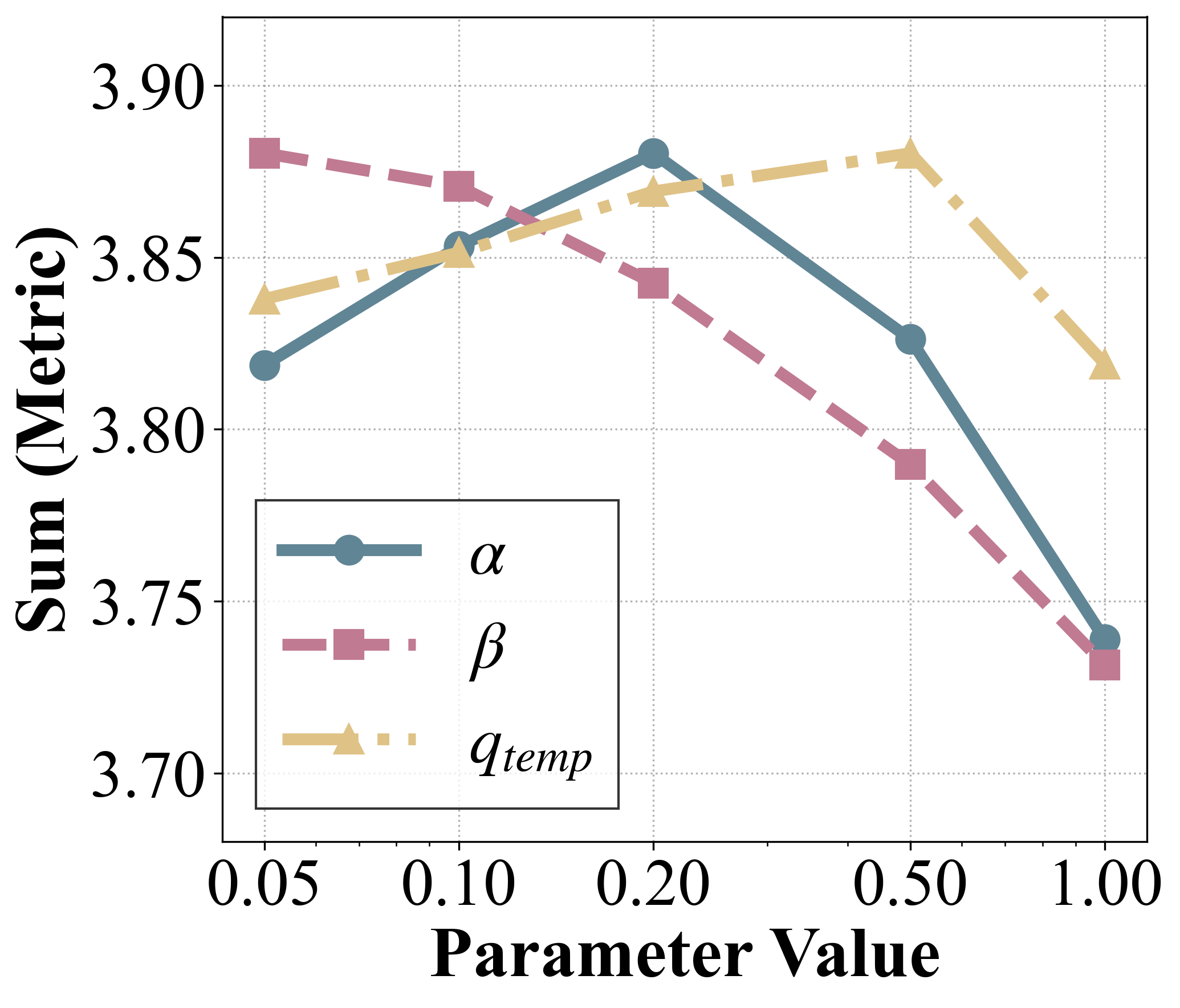}
      \caption{Hyperparameter Study on Beauty.}
      \label{fig 4}
   \end{minipage}
\end{figure}

The results are shown in Figure \ref{fig 3}, where the number of ties is omitted since it is small. Across the three datasets, the high-confidence augmented view has higher target co-occurrence counts than the low-confidence view in about 65\%–69\% of cases, which deviates substantially from a random distribution. This suggests that the operation confidence of co-augmentation is a reliable proxy for view quality. In a small fraction of cases, the low-confidence view receives a higher quality score, which may be due to the item popularity bias: a few popular items may increase target co-occurrence counts even when the view is less confidence. We also find from individual examples that the confidence scores of the two views are very close (e.g., 0.78 and 0.77), in which case the lower-confidence view may still obtain a higher quality score. Overall, the results indicate that higher confidence is statistically associated with higher-quality views, but not deterministically. This motivates our adaptive quality-aware weighting, rather than the hard filtering or manually fixed weighting used in prior works.

\subsection{Robustness Analysis}

\begin{table*}[ht]
    \renewcommand{\arraystretch}{0.8}
    \caption{Robustness analysis. MR and NG are short for MRR and NDCG. The table shows the performance comparison of different models on the simulated test sets. The best performance is denoted in bold fonts.}
    \label{tab4}
    \scalebox{0.9}{
    \begin{tabular}{c|c| ccc ccc ccc |c|c|c}
         \toprule
         Dataset & Model & HR@5 & HR@10 & HR@20 & MR@5 & MR@10 & MR@20 & NG@5 & NG@10 & NG@20 & Sum & Raw & dist \\
         \midrule

         \multirow{6}{*}{\makecell[c]{Simulated \\Beauty}} 
         &QCMP-CL &\textbf{0.4469} &\textbf{0.5477} &\textbf{0.6624} &\textbf{0.3109} &\textbf{0.3244} &\textbf{0.3323} &\textbf{0.3448} &\textbf{0.3774}&\textbf{0.4063} &\textbf{3.7535} &\textbf{3.8803} &-3.26\% \\
         &TCLARec &0.4306 &0.5276 &0.6424 &0.3057 &0.3186 &0.3265 &0.3368 &0.3681 &0.3971 &3.6540 &3.7740 &\textbf{-3.17\%} \\
         &ICSRec &0.4045 &0.5100 &0.6278 &0.2709 &0.2849 &0.2931 &0.3041 &0.3381 &0.3679 &3.4013 &3.5748 &-4.85\% \\
         &STEAM &0.4081 &0.5080 &0.6279 &0.2718 &0.2851 &0.2933 &0.3057 &0.3379 &0.3681 &3.4062 &3.5561 &-4.21\% \\
         &CoSeRec &0.3925 &0.5020 &0.6262 &0.2567 &0.2714 &0.2799 &0.2905 &0.3259 &0.3572 &3.3022 &3.4529 &-4.36\% \\
         &CL4SRec &0.3911 &0.4882 &0.6008 &0.2690 &0.2819 &0.2896 &0.2993 &0.3307 &0.3590 &3.3100 &3.4253 &-3.36\% \\
         &SASRec &0.3502 &0.4373 &0.5550 &0.2422 &0.2537 &0.2617 &0.2690 &0.2970 &0.3266 &2.9933 &3.1868 &-6.07\% \\
         \midrule
         \multirow{6}{*}{\makecell[c]{Simulated \\Yelp}} 
         &QCMP-CL &\textbf{0.6774} &\textbf{0.8544} &0.9546 &\textbf{0.4376} &\textbf{0.4615} &\textbf{0.4688} &\textbf{0.4971} &\textbf{0.5546} &\textbf{0.5804} &\textbf{5.4868} &\textbf{5.6661} &\textbf{-3.16\%}\\
         &TCLARec &0.6536 &0.8357 &0.9514 &0.4193 &0.4437 &0.4522 &0.4774 &0.5364 &0.5662 &5.3365 &5.5523 &-3.88\% \\
         &ICSRec &0.6183 &0.8092 &\textbf{0.9551} &0.3875 &0.4129 &0.4235 &0.4447 &0.5064 &0.5438 &5.1014 &5.3993 &-5.51\% \\
         &STEAM &0.6328 &0.8160 &0.9355 &0.3991 &0.4237 &0.4324 &0.4571 &0.5165 &0.5472 &5.1606 &5.4176 &-4.74\% \\
         &CoSeRec &0.6148 &0.7966 &0.9293 &0.3853 &0.4097 &0.4192 &0.4421 &0.5011 &0.5350 &5.0331 &5.2734 &-4.56\% \\
         &CL4SRec &0.6003 &0.7910 &0.9280 &0.3721 &0.3977 &0.4077 &0.4286 &0.4905 &0.5257 &4.9419 &5.1849 &-4.68\% \\
         &SASRec &0.5352 &0.7334 &0.8848 &0.3135 &0.3400 &0.3508 &0.3683 &0.4325 &0.4711 &4.4301 &4.8109 &-7.91\%  \\ 
         \midrule
         \multirow{6}{*}{\makecell[c]{Simulated \\Sports}} 
         &QCMP-CL&\textbf{0.4337} &\textbf{0.5599} &\textbf{0.6950} &\textbf{0.2857} &\textbf{0.3025} &\textbf{0.3118} &\textbf{0.3224} &\textbf{0.3632} &\textbf{0.3973} &\textbf{3.6720} &\textbf{3.8251} &\textbf{-4.00\%} \\
         &TCLARec &0.4186 &0.5440 &0.6803 &0.2740 &0.2906 &0.3000 &0.3099 &0.3503 &0.3847 &3.5527 &3.7122 &-4.29\% \\
         &ICSRec &0.3988 &0.5246 &0.6701 &0.2552 &0.2719 &0.2820 &0.2908 &0.3314 &0.3682 &3.3931 &3.5668 &-4.87\% \\
         &STEAM &0.3929 &0.5259 &0.6724 &0.2475 &0.2652 &0.2753 &0.2836 &0.3265 &0.3635 &3.3530 &3.5476 &-5.48\% \\
         &CoSeRec &0.3813 &0.5019 &0.6377 &0.2420 &0.2580 &0.2674 &0.2765 &0.3154 &0.3497 &3.2300 &3.4559 &-6.53\% \\
         &CL4SRec &0.3648 &0.4854 &0.6251 &0.2404 &0.2563 &0.2659 &0.2712 &0.3100 &0.3452 &3.1647 &3.3805 &-6.38\% \\
         &SASRec &0.3155 &0.4275 &0.5688 &0.1987 &0.2135 &0.2232 &0.2276 &0.2636 &0.2993 &2.7383 &2.9622 &-7.56\% \\ 
         \bottomrule
    \end{tabular}
    }
\end{table*}

To evaluate the model’s generalization and noise robustness under noisy interactions and sequence perturbations, we construct simulated test sets. Specifically, for each sequence in the original test set, we apply item-wise keep, delete, and insert operations with a ratio of 4:3:3, which simulates missing interactions and noisy exposures commonly observed in real-world recommender systems \cite{30-}. We train all models on the original training set and evaluate them on the simulated test set. Table \ref{tab4} reports the performance of different models on three simulated test sets, where Sum denotes the sum of the nine metrics and Raw is the sum of the metrics reported on the clean test sets (Table \ref{tab2}). We define $dist=(Sum-Raw)/Raw$ to quantify the overall performance degradation.

We have the following observations. First, sequence perturbations significantly degrade the performance of all models. The drop is most pronounced for the purely self-supervised recommendation method, indicating that representations learned only from next-item supervision are more sensitive to noisy interactions. Second, models that incorporate contrastive learning and stronger view construction are generally more robust. Heuristic-augmentation methods such as CL4SRec and CoSeRec perform well on Beauty and Yelp, but still suffer a large drop on Sports, indicating the instability of random augmentation against strong noise. In comparison, TCLARec reduces dist to -3.17\%/-3.88\%/-4.29\% on the three datasets, highlighting the advantage of learnable sequence augmentation against insertion/deletion noise. STEAM also shows strong robustness, because its self-supervised correction task can mitigate inherent noisy interactions in the raw sequences.

QCMP-CL achieves the best or near-best overall results on all three simulated test sets, and it is advantageous in both absolute performance and degradation. In terms of absolute performance, QCMP-CL ranks first on most metrics, and is only slightly below ICSRec on HR@20 for Simulated Yelp. In terms of degradation, QCMP-CL gets the smallest drop on Yelp and Sports, outperforming TCLARec and other baselines. On Beauty, QCMP-CL shows a degradation of -3.26\%, close to TCLARec’s -3.17\%, while still achieving higher absolute performance. These results align with our design intuition. On the one hand, the collaborative augmentation module is pre-trained in Stage I via a ``corrupt-and-reconstruct" task, learning to preserve key items, remove noise, and insert auxiliary items under insertion/deletion perturbations, which improves stability under corruption when testing. On the other hand, the ``collaborative multi-view + quality weight" design in Stage II further suppresses incorrect alignments caused by low-quality views, making the learned sequence representations less sensitive to random noise. Overall, QCMP-CL not only performs best on clean test sets, but also maintains stronger performance with smaller degradation under heavily perturbed, noisy test scenarios.

\subsection{Hyperparameter Study}

The weight $\alpha$ for $L_{qmp}$, the weight $\beta$ for $L_{cl}$, and the temperature parameter $q_{temp}$ in Eq. \ref{eq15} are key hyperparameters. We vary each of them in $[0.05,0.1,0.2,0.5,1.0]$ and examine the performance trends on the Beauty dataset.

As shown in Figure \ref{fig 4}, we have the following observations. $\alpha$ controls the contribution of $L_{qmp}$, and the best overall performance is achieved at $\alpha=0.2$. When $\alpha$ is too small, the quality-aware multi-positive contrastive signal becomes insufficient; when it is too large, optimization is overly shifted from next-item prediction to representation alignment. Both cases lead to a clear performance drop. In contrast, $\beta$ should be set smaller, since $L_{cl}$ serves as an auxiliary regularizer. A slight increase in $\beta$ has limited impact, but an overly large $\beta$ harms both the main contrastive objective and the next-item prediction task, substantially degrading performance. Moreover, $q_{temp}$ determines how ``sharp" the view-weight distribution is. A too-small $q_{temp}$ makes the weights overly peaked, reducing the utilization of complementary information across views; a too-large $q_{temp}$ makes the weights close to uniform, weakening the quality-aware mechanism. Overall, setting $q_{temp}=0.5$ balances complementarity and quality discrimination, yielding robust gains.

\section{Conclusion}

This work proposes QCMP-CL, a contrastive learning-based model for sequential recommendation, aiming to provide diverse and high-quality contrastive views for contrastive learning. QCMP-CL first pre-trains a collaborative augmentation module that generates two positive views under two complementary collaborative contexts. We then design a quality-aware multi-positive contrastive objective that assigns adaptive weights to the positive views based on the confidence of augmentation operations, thereby controlling their supervision strength and mitigating the false-positive issue. Extensive experiments on three datasets show that QCMP-CL consistently outperforms strong baselines in both overall performance and robustness, and additional ablation and statistical analyses further validate the effectiveness of each component.

\section{Limitations and Future Work}
Despite the improvements of QCMP-CL, several limitations remain. First, although the confidence of augmentation operation is an effective proxy, it is not always perfectly interpretable as semantic view quality, and it may be sensitive to the pre-training level of the augmentation module. Second, our contrastive learning still relies on in-batch negatives without explicitly assessing their reliability, where false negatives may interfere with representation learning. Third, although freezing the augmentation module stabilizes two-stage training, it may limit the end-to-end adaptation of augmentation strategies to downstream recommendation objectives. Finally, compared with heuristic augmentation, our collaborative multi-view construction incurs additional computation.

To address these limitations, we will extend QCMP-CL along four directions. First, for positive-quality modeling, we plan to integrate multiple quality signals (e.g., representation consistency between views and anchors) to build a more robust and interpretable quality estimator. Second, on the negative side, we will explore a dual-sided quality-aware contrastive framework that down-weights potential false negatives using similarity thresholds or nearest-neighbor retrieval. Third, we will fine-tune a small subset of augmentation module parameters jointly with the recommender to improve the adaptation to recommendation objectives. Finally, we will consider selective view sampling and lightweight encoders to reduce redundant computation.


\bibliographystyle{ACM-Reference-Format}
\balance
\bibliography{reference}

\end{sloppypar}
\end{document}